\documentclass[a4paper,11pt]{article}
\usepackage{pos}
 
\usepackage{amsmath}
\usepackage{siunitx}
\usepackage{todonotes}
\usepackage{physics}
\usepackage{cleveref}

\newcommand{\eff}{\text{eff}}
\newcommand{\zth}{^{(0)}}
\newcommand{\fst}{^{(1)}}
\newcommand{\qcdiso}{$\text{QCD}_{\text{iso}}$}

\title{Isospin-breaking Effects in Octet and Decuplet Baryon Masses}

\author*[a]{Alexander M. Segner}
\author[b]{Andrew D. Hanlon}
\author[a]{Renwick J. Hudspith}
\author[c]{Andreas Risch}
\author[a,d,e,f]{Hartmut Wittig}

\affiliation[a]{Institut für Kernphysik, Johannes Gutenberg-Universität, \\
    Johann-Joachim-Becher-Weg 45, 55128 Mainz, Germany}

\affiliation[b]{Physics Department, Brookhaven National Laboratory, \\
    Upton, New York 11973, USA}

\affiliation[c]{John von Neumann-Institut für Computing NIC, Deutsches Elektronen-Synchrotron DESY, \\
    Platanenallee 6, 15738 Zeuthen, Germany}

\affiliation[d]{Helmholtz Institut Mainz, \\
    Staudingerweg 18, 55128 Mainz, Germany}

\affiliation[e]{Helmholtzzentrum für Schwerionenforschung, \\ 
    64291 Darmstadt, Germany}

\affiliation[f]{PRISMA$^{+}$ Cluster of Excellence, Johannes-Gutenberg-Universit{\"a}t Mainz,\\
    Staudingerweg 9, 55128 Mainz, Germany}

\emailAdd{alsegner@uni-mainz.de}
\emailAdd{ahanlon@uni-mainz.de}
\emailAdd{renwick.james.hudspith@googlemail.com}
\emailAdd{andreas.risch@uni-mainz.de}
\emailAdd{hartmut.wittig@uni-mainz.de}

\abstract{We present work designed to compute baryon masses on $N_f = 2 + 1$ CLS ensembles including
isospin-breaking effects due to non-degenerate light quark masses and electromagnetic interactions.
These effects are determined at leading order via a perturbative expansion around the iso-symmetric
theory. We furthermore apply a group-theoretical operator construction for the various interpolators
describing the different members of the baryon octet and decuplet based on a classification by spin,
parity, and flavor content.}

\FullConference{%
 The 38th International Symposium on Lattice Field Theory, LATTICE2021
  26th-30th July, 2021
  Zoom/Gather@Massachusetts Institute of Technology
}

\begin{document}

\maketitle

\section{Introduction}\label{sec:introduction}

We are in an era of precision lattice QCD physics, where contributions from QED and strong-isospin-breaking can no longer be ignored. 
An example where these contributions are of significant importance is in the lattice determination of the anomalous magnetic moment of the muon $(g-2)_\mu$. 
Commonly, one of the largest uncertainties in any lattice determination of $(g-2)_\mu$ comes from the scale setting and it is of great importance that a calculation has the physical scale set accurately (below $1\%$) incorporating QED and isospin-breaking effects

As ensembles for lattice QCD are often generated for the isosymmetric theory \qcdiso, perturbative methods were developed to incorporate isospin-breaking effects into calculations on these isosymmetric ensembles \cite{deDivitiis:2011eh,deDivitiis:2013xla}.
The goal of this project is to reduce the contribution of isospin-breaking effects to the uncertainty of the lattice scale $a$ for $N_f=2+1$ CLS ensembles. 

In ref. \cite{Bruno:2016plf}, the lattice scale for the of CLS ensembles was determined using a combination of pion and kaon decay constants.
While the final result has a total error at the level of 1\%, the incorporation of isospin-breaking corrections turns out to be quite difficult \cite{Carrasco:2015xwa}.
In this project, we investigate the prospects of precision scale setting using the masses of the lowest-lying baryon octet and decuplet, for which isospin-breaking effects are simpler to incorporate.

\section{Inclusion of Perturbative Isospin-breaking Effects by Reweighting}\label{sec:ib}

The procedure for the calculation of isospin-breaking corrections to the hadron masses follows the perturbative approach introduced by the RM123 collaboration \cite{deDivitiis:2011eh,deDivitiis:2013xla}.
Here, the QCD+QED action $S$ described by the set of parameters $\varepsilon=(\beta,e^2,m_u,m_d,m_s)$ (the inverse QCD coupling, the squared QED coupling, and the masses of the up, down, and strange quarks) is expanded around the isospin-symmetric action $S^{(0)}$ with parameters $\varepsilon^{(0)}=(\beta^{(0)},0,m_{ud}^{(0)},m_{ud}^{(0)},m_s^{(0)})$ in terms of these parameters.
$S^{(0)}[U,\psi,\bar\psi]=S_g\zth[U]+S_q\zth[U,\psi,\bar\psi]$ here denotes an isospin-symmetric action consisting of the Lüscher-Weisz action $S_g\zth$ for the gauge fields and an $O(a)$ improved action $S_q\zth$ for Wilson fermions with $N_f=2+1$ flavors \cite{Bruno:2014jqa}. In this work we use gauge ensembles simulated by the CLS
collaboration.
The QCD+QED action can be divided into three parts, a QCD gauge action, a QED gauge action, and a quark action:
\[
    S[U,A,\psi,\bar\psi]=S_g[U]+S_\gamma[A]+S_q[U,A,\psi,\bar\psi]
\]
For the QED gauge action in finite volume, we use the non-compact QED$_{\text{L}}$ prescription \cite{Hayakawa:2008an} in Coulomb gauge, which introduces an infrared regularisation by eliminating the zero-momentum modes of the photon field by setting
\[
    \sum_{\vb x}A^{\vb x,t}\equiv0
\]
on every timeslice $t$.

In this setup, expectation values for operators in the full theory, i.e.
\begin{align}
\begin{aligned}
    \ev{\mathcal O}_S=&\frac1{Z}\int\mathcal DU\mathcal DA\mathcal D\psi\mathcal D\bar\psi\,\mathcal O[U,A,\psi,\bar\psi]e^{-S[U,A,\psi,\bar\psi]} \\
    =&\frac1Z\int\mathcal DU
    \underbrace{\qty(\frac1{Z_{q\gamma}[U]}\int\mathcal DA\mathcal D\psi\mathcal D\bar\psi\,\mathcal O[U,A,\psi,\bar\psi]e^{-S_\gamma[A]-S_q[U,A,\psi,\bar\psi]})}_{=:\ev{\mathcal O}_{S_{q\gamma}}[U]}
    \underbrace{Z_{q\gamma}[U]e^{-S_g[U]}}_{=:\exp(-S_\eff[U])}
\end{aligned}\label{eq:ev}
\end{align}
can be expressed in terms of those in isosymmetric QCD by reweighting~\cite{Risch:2017xxe,Risch:2018ozp}:
\begin{align}
\begin{aligned}
    \ev{\mathcal O}_{S}=&\frac{\frac1{Z_\eff\zth}\int\mathcal DU\ev{\mathcal O}_{S_{q\gamma}}[U]\frac{\exp(-S_\eff[U])}{\exp(-S_\eff\zth[U])}e^{-S_\eff\zth[U]}}{\frac1{Z_\eff\zth}\int\mathcal DU\,\frac{\exp(-S_\eff[U])}{\exp(-S_\eff\zth[U])}e^{-S_\eff\zth[U]}}=\frac{\ev{R\ev{\mathcal O}_{S_{q\gamma}}}_{S_\eff\zth}}{\ev R_{S_\eff\zth}} 
\end{aligned}\label{eq:reweighting} ,
\end{align}
where $R$ denotes the reweighting factor
\begin{align}
    R=\frac{\exp(-S_\eff)}{\exp(-S_\eff^{(0)})}=\frac{\exp(-S_g)\,Z_{q\gamma}}{\exp(-S_g^{(0)})\,Z_q^{(0)}},
\end{align}
which replaces the Boltzmann weight associated with the effective action of QCD$_{\mathrm{iso}}$ by its counterpart in QCD+QED.
The effective actions are defined as
\begin{align*}
    S_\eff[U]=S_g[U]-Z_{q\gamma}[U]=&S_g[U]-\log(\int\mathcal DA\mathcal D\psi\mathcal D\bar\psi\,e^{-S_\gamma[A]-S_q[U,A,\psi,\bar\psi]}) \\
    S_\eff\zth=S_g\zth[U]-Z_q\zth[U]=&S_g\zth[U]-\log(\int\mathcal D\psi\mathcal D\bar\psi\,e^{-S_q\zth[U,\psi,\bar\psi]}).
\end{align*}

To evaluate the expectation value $\ev{\mathcal O}_{\mathrm{q\gamma}}$ and the reweighting factor $R$ in \cref{eq:reweighting} we use perturbation theory and expand the latter expressions in terms of the parameters
\begin{align*}
\Delta\varepsilon=&\varepsilon-\varepsilon^{(0)} = (\Delta\beta,e^2,\Delta m_u,\Delta m_d,\Delta m_s) \\
=&(\beta-\beta\zth,e^2,m_u-m_{ud}^{(0)},m_d-m_{ud}^{(0)},m_s-m_s\zth)
\end{align*}
around $\varepsilon^{(0)}$. Operators that depend on the QED gauge links $\exp(\mathrm{i}aeQA)$, where $Q$ denotes the matrix of quark charges, also have to be expanded in $e$, i.e. $\mathcal{O}=\mathcal{O}^{(0)}+e\mathcal{O}^{(\frac{1}{2})}+\frac{1}{2}e^{2}\mathcal{O}^{(1)}+O(e^{3})$.

\section{Baryon Operators}\label{sec:operators}

The operator bases used for the various baryons considered in this work were first introduced in Ref.~\cite{Basak:2005ir}. 
These operators are obtained from a group-theoretical construction of the coefficients $\lambda$ for a baryonic interpolator
\[
    \mathcal O_B=\sum_{a,b,c,f_i,\mu_j}\varepsilon_{abc}\lambda^{\mu_1,\mu_2,\mu_3}_{f_1,f_2,f_3}q_{\mu_1}^{f_1,a}q_{\mu_2}^{f_2,b}q_{\mu_3}^{f_3,c},
\]
where $f_i\in{u,d,s}$ label the flavor of the quarks, $a,b,c$ are color indices, and $\mu_i$ are Dirac-spinor indices. 
In this project we use Wuppertal-smeared fields, i.e. $q=W\Psi$, where $\Psi$ is a point source and $W$ is an APE-smeared smearing operator.

This construction distinguishes different baryons by their symmetries w.r.t the quark flavors $f_i$ and the flavors themselves.
For each flavor symmetry, a set of operators based on the third spin component and parity are constructed from a tensor product of Weyl spinors.
The only differences between the operators we use and those in \cite{Basak:2005ir} are their normalization and that we do not make use of the totally symmetric $\Sigma$ and $\Xi$ operators.
Furthermore, we do not use the H-irreps of the nucleon and (in the isosymmetric case) $\Lambda$, as well as the $G_1$-irreps of the $\Delta$ and $\Omega$ states since we are only interested in the ground state energies.

These operators are constructed in the Dirac-Pauli basis, in which the parity operator only acts on the first two indices of a Dirac spinor and the spin-z operator only acts on the last two spinor indices, which allows for a convenient construction of the operators based on parity and spin-z eigenvalues in terms of Weyl spinors $\chi,\xi$ such that the Dirac spinor is given by $\psi=\chi\otimes\xi$:
\begin{align*}
    \mathcal P\psi(\vb x,t)=&\gamma_0\psi(-\vb x,t)=((\sigma_3\chi)\otimes\xi)(-\vb x,t) , \\ 
    S_z\psi(\vb x,t)=&-i\gamma_1\gamma_2\psi(\vb x,t)=\chi\otimes(\sigma_3\xi)(\vb x,t) .
\end{align*}
Using this fact, baryonic operators are constructed from three Weyl spinors on which the parity operator acts and three Weyl spinors defining the spin of the baryon.
The tensor product of different symmetrizations of these combinations of three Weyl spinors then define the baryon operators. 
Since the baryon operators are already antisymmetric w.r.t their color indices, the combined symmetry of spinor and flavor indices has to be chosen such that the operators are antisymmetric under the exchange of two quarks.
As the different baryons are classified according to their flavor symmetries, the Dirac indices thus have to make the operators symmetric or mixed-symmetric under simultaneous exchange of flavor and spin indices.

In total, this procedure results in 116 different operators: 58 for each parity eigenvalue.
Because more than one of these operators are expected to have overlap with the same ground state, this allows us to perform a GEVP in order to better control the excited states \cite{Blossier:2008tx,Blossier:2009kd}.
The sizes of the correlator matrices for the different baryons are listed in \cref{tab:corr_sizes}, with each baryon having one correlator matrix for each spin-z eigenvalue.

\begin{table}[htb]
\centering
\caption{Sizes of the correlator matrices for each particle.}
\label{tab:corr_sizes}
\begin{tabular}{r|c|c|c|c|c|c|c}
    & $N$ & $\Lambda$ & $\Sigma/\Xi$ & $\Delta/\Omega$ & $\Sigma^*/\Xi^*$ & $\Sigma$-$\Lambda$-mixing & $\Sigma^*$-$\Lambda$-mixing \\
    \hline
    Correlator size & $3\times3$ & $4\times4$ & $3\times3$ & $2\times2$ & $2\times2$ & $7\times7$ & $2\times2$ 
\end{tabular}
\end{table}

\section{Baryonic two-point functions}\label{sec:bar2pt}

As we are aiming for the determination of baryon masses, we consider baryonic two-point functions consisting of zero-momentum projected baryonic creation and annihilations operators $\overline{\mathcal{B}}=\overline{B}\overline{\Psi}\overline{\Psi}\overline{\Psi}$ and $\mathcal{B}=B\Psi\Psi\Psi$. 
In the following, we assume that $\overline{\mathcal{B}}$ and $\mathcal{B}$ do not depend on $e$, i.e. $\overline{\mathcal{B}}=\overline{\mathcal{B}}{}^{(0)}$ and $\mathcal{B}=\mathcal{B}^{(0)}$. 
In particular, we may apply QCD-covariant but not QCD+QED-covariant operator smearing. 
The latter would lead to contributions from $\overline{\mathcal{B}}{}^{(\frac{1}{2})}$, $\overline{\mathcal{B}}{}^{(1)}$, $\mathcal{B}^{(\frac{1}{2})}$ and $\mathcal{B}^{(1)}$ and hence to additional diagrams.

\begin{figure}[htb]
\begin{align*}
\langle\mathcal{B}\overline{\mathcal{B}}\rangle_S &=
\begin{aligned}[t]
\Big\langle&\vcenter{\hbox{\includegraphics[width=3cm]{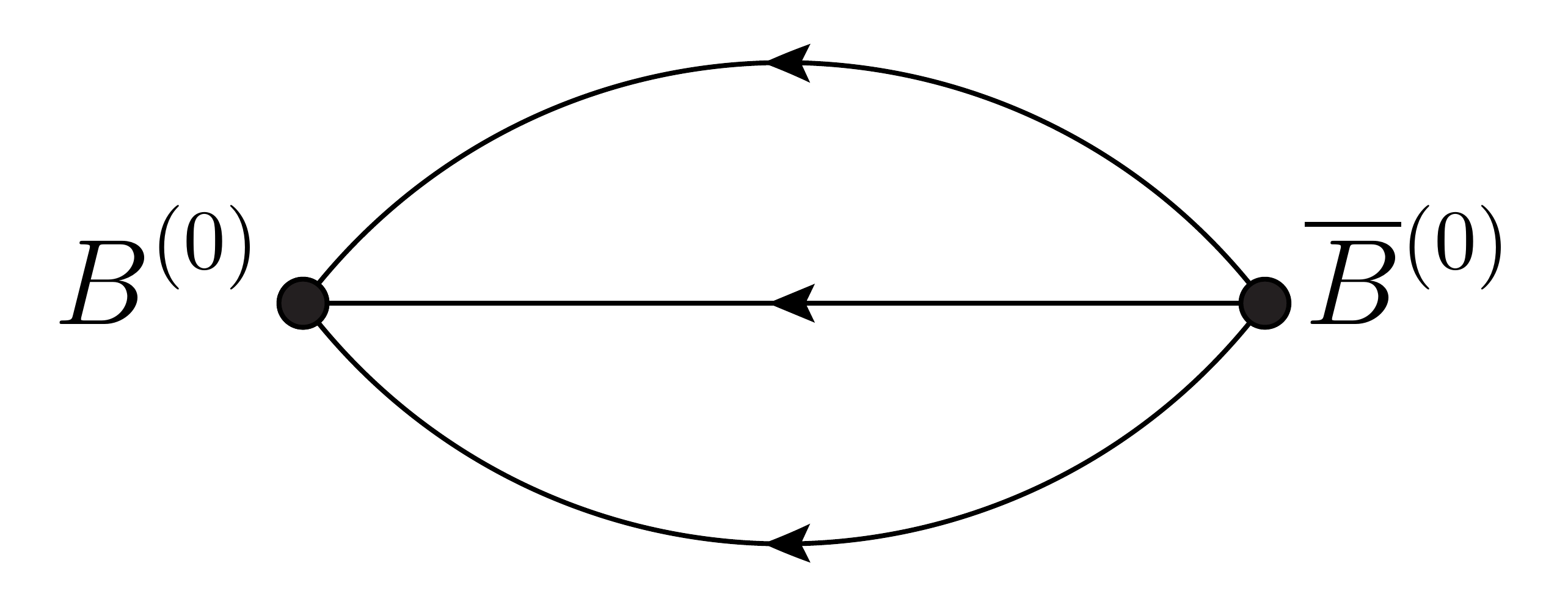}}}+\sum_f\Delta m_f\vcenter{\hbox{\includegraphics[width=3cm]{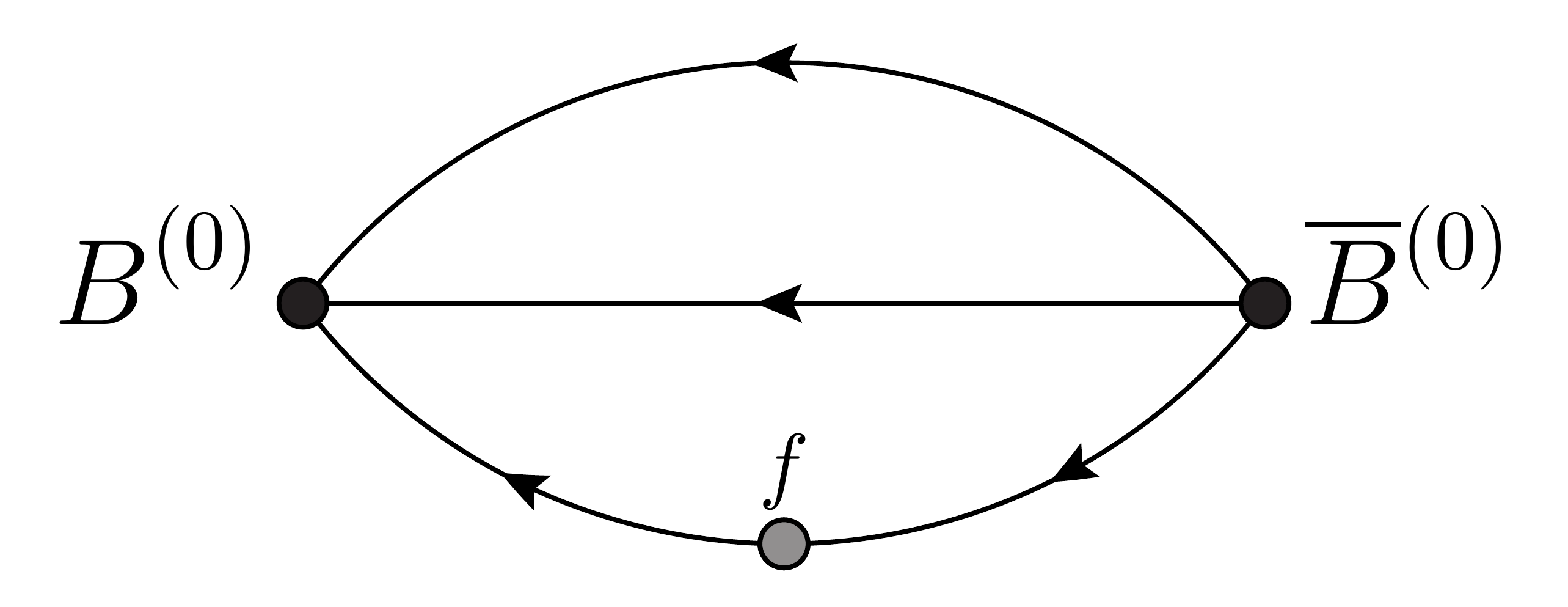}}}  \\
&+ e^2\qty(\vcenter{\hbox{\includegraphics[width=3cm]{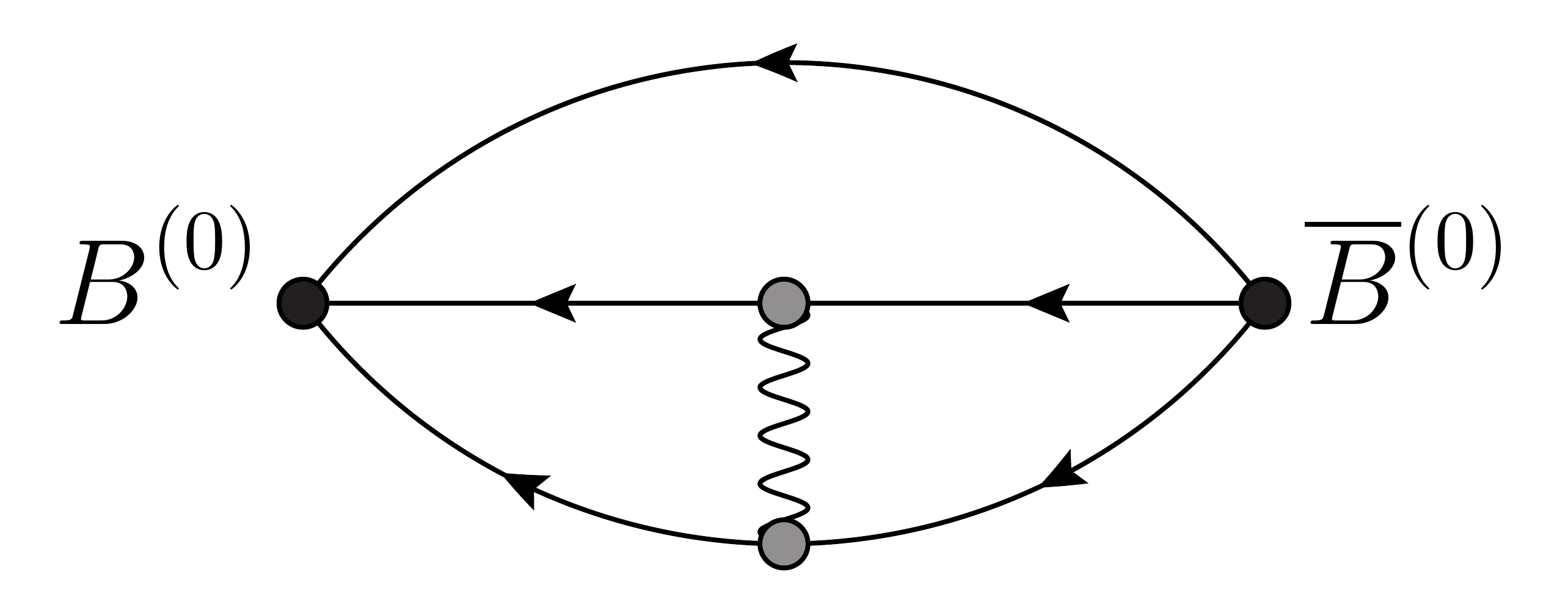}}} + \vcenter{\hbox{\includegraphics[width=3cm]{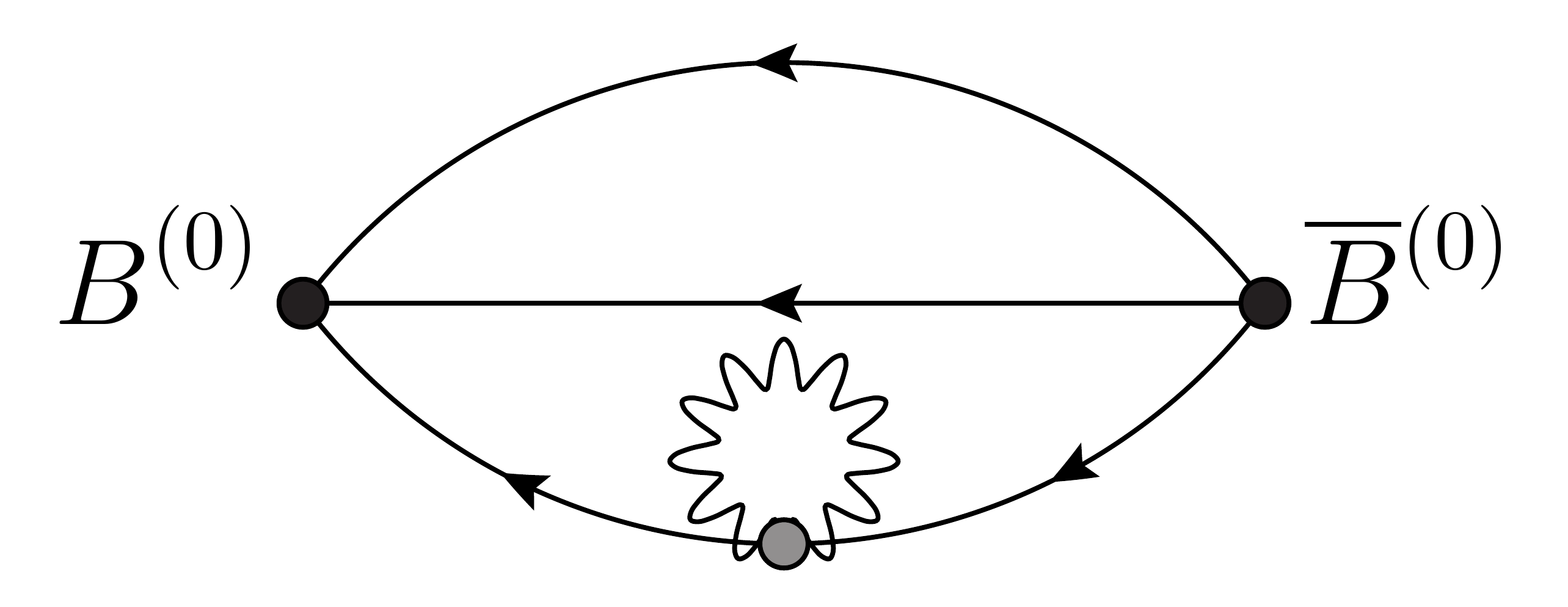}}} + \vcenter{\hbox{\includegraphics[width=3cm]{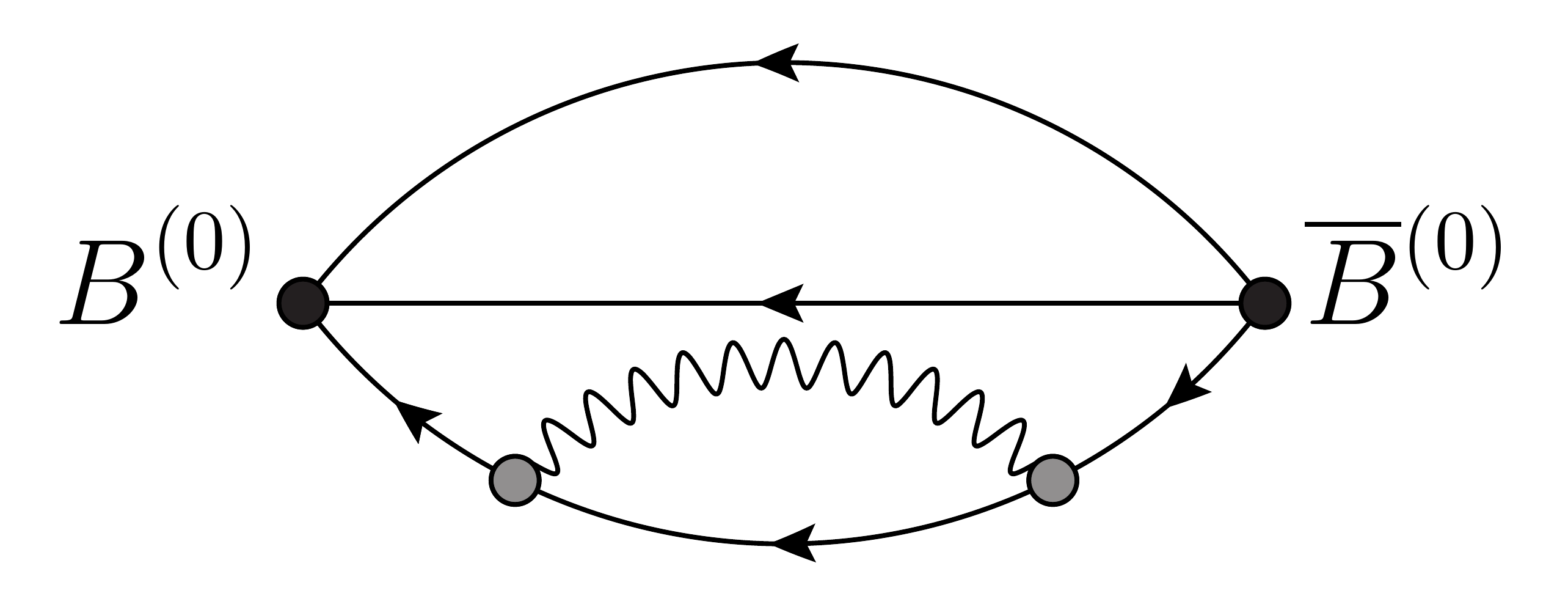}}}) \\
    &+\cdots\Big\rangle_{S_\eff^{(0)}}
\end{aligned}
\end{align*}
\caption{Diagrammatic expansion of QCD+QED correlation functions in terms of isospin-breaking parameters around correlators in \qcdiso. The index $f$ counts the different quark flavors.}
\label{fig:diagrams}
\end{figure}

As an example, the expansion of baryonic correlation functions is shown in \cref{fig:diagrams}, in which sequential propagators are used in the first-order terms.
The full expansion also contains diagrams with quark-disconnected parts and corrections acting in the sea quark sector.
However, in this project we neglect isospin-breaking effects in the sea quarks (for the time being) and we set $\Delta\beta$ to 0.
As a consequence, only quark-connected contributions remain and $R=1$ in \cref{eq:reweighting}.
Nonetheless, we calculate the diagrams with a photon line connected to one of the valence quarks in case we decide to include the disconnected contributions at a later stage since they are needed for diagrams in which a sea quark electromagnetically interacts with a valence quark.
Note that in \cref{sec:optimizations} only contributions shown in \cref{fig:diagrams} are discussed. 
However, the additional diagrams can trivially be incorporated into the optimizations described, which moreover remove any significant overhead arising from the additional computation of these diagrams.

In order to extract the mass of the lowest  state incorporated in a baryonic two-point function, we consider the asymptotic time dependence $C(t_2,t_1)= c e^{-m(t_2-t_1)}$. 
Expanding the latter in terms of the isospin-breaking parameters $X=X\zth+\sum_i\Delta\varepsilon_iX_i+\mathcal O(\Delta\varepsilon^2)$ with $X\in\{c,m,C\}$, one finds the zeroth- and first-order contributions
\begin{align*}
C\zth(t_2,t_1)=&c^{(0)}e^{-m^{(0)}(t_2-t_1)}, \\
C_i\fst(t_2,t_1)=&\qty(c^{(1)}_i-c^{(0)}m^{(1)}_i(t_2-t_1))e^{-m^{(0)}(t_2-t_1)}.
\end{align*}
$m\zth$ can then be reconstructed from the usual definition of the effective mass
\[
    (am_\eff)\zth(t_{2},t_{1}):=\log\frac{C\zth(t_2,t_1)}{C\zth(t_2+a,t_1)},
\]
with the analogous definition for the first-order terms
\[
    (am_\eff)_{i}\fst(t_{2},t_{1}):=\frac{C_i\fst(t_2,t_1)}{C\zth(t_2,t_1)}-\frac{C_i\fst(t_2+a,t_1)}{C\zth(t_2+a,t_1)}.
\]

\section{Computational Optimizations}\label{sec:optimizations}

We compute correlation functions using the operators described in \cref{sec:operators} with a program written in \texttt{C++} making use of the libraries \texttt{OpenQCD} for quark propagator inversions and \texttt{QDP++} for the remaining computations. For better statistics for the same amount of computing time, we apply all-mode averaging\ \cite{Shintani:2014vja}.

As the chosen set of operators described in \cref{sec:operators} yields a large number (528) of non-vanishing correlator matrix entries, the contractions can easily become a bottleneck in the computation of the different correlators, especially in the case of isospin-breaking corrections.
Therefore, a number of optimizations have been applied to the production code in order to keep the computational costs small compared to the inversions.

First, the contractions are simplified algebraically by identifying the up and down quark propagators with the light quark $l$ used in the simulations, which is possible as we use isospin-symmetric ensembles.
A similar simplification is done for up and down propagators including a photon vertex:
In the contractions, these terms come with a factor $e_fe$ for each three-point-vertex or $(e_fe)^2$ in the case of a four-point-vertex, where $e$ is the electromagnetic coupling and 
\[
    e_f=\begin{cases} \frac23 & f=u \\ -\frac13 & f\in\{d,s\}\end{cases}. 
\]
Thus, in a diagram with photon interactions, the product of the fractional charges of the quarks at each photon vertex yields simply a prefactor for the contractions with $S_d$ and $S_u$ replaced by $S_l$.
Note, that the charge multiplicity $e_f$ is already absorbed into the definition of the vertices in \cref{fig:diagrams}.

This leaves 101580 individual, color-contracted terms of the form:
\begin{align}
    T_{\mu_1\mu_2\mu_3}^{f_1f_2f_3}=\sum\limits_{\smqty{a,b,c \\ a',b',c'}}\varepsilon_{abc}\varepsilon_{a'b'c'}\lambda_{f_1f_2f_3}^{\mu_1\mu_2\mu_3}\lambda_{f_1f_2f_3}^{\mu_4\mu_5\mu_6}S^{f_1,aa'}_{\mu_1\mu_4}S^{f_2,bb'}_{\mu_2\mu_5}S^{f_3,cc'}_{\mu_3\mu_6}, \label{eq:terms}
\end{align}
which divide into 8304 isosymmetric contributions, 38316 contributions from mass detuning, and 54960 from QED corrections.

However, many of these terms appear in multiple contractions (possibly differing in the coefficients $\lambda$), enabling a reduction of the computational cost by reusing these terms after their first computation.
This reduces the number of terms to be computed to 10104 unique terms of the form
\[
    \tilde T_{\mu_1\mu_2\mu_3}^{f_1f_2f_3}=\sum\limits_{\smqty{a,b,c \\ a',b',c'}}\varepsilon_{abc}\varepsilon_{a'b'c'}S^{f_1,aa'}_{\mu_1\mu_4}S^{f_2,bb'}_{\mu_2\mu_5}S^{f_3,cc'}_{\mu_3\mu_6}
\]
such that all correlators can be expressed as linear combinations of these unique terms.
This, however, leads to the next challenge: since each term resembles a complex field on the lattice, this is still an unreasonable amount of data to hold in memory at once, especially on large ensembles.
This problem can be circumvented by making use of the fact that each correlator has a fixed combination of flavor indices $(f_1,f_2,f_3)$, where we count the sequential propagators from isospin-breaking contributions as separate flavors so that there are a total of 25 different flavor combinations.
One can therefore split the set of correlators into much smaller subsets that can be computed in one go without the need of keeping the terms in memory for any other subset.
These small subsets then contain at most 640 unique terms, which is much more manageable.

Further optimizations were necessary due to the nature of the \texttt{QDP++} library used for the computation of the contractions.
This library is optimized for matrix-based computations on the lattice, but has rather slow routines for retrieving data based on color or spin indices.
To avoid these routines, propagators are saved as \texttt{std::vector<std::vector<LatticeColorMatrix>>}, so that \texttt{QDP++}'s \texttt{peekSpin}-routine only needs to be called 16 times per propagator in order to store the propagator in this format.

Once all propagators for a given flavor combination are calculated, a lookup table for all color-contracted terms is constructed for that flavor combination.
This lookup table is finally used to calculate the contractions.
It was found to be beneficial, performance wise, to use precompiled functions that return the complete contraction as a single expression in terms of the elements of the lookup table.
An example for such a function calculating the contraction for a simple isospin-symmetric nucleon correlator would take references to a \texttt{std::map<std::tuple<int, int, int, int, int, int>, LatticeComplex> map} and a \texttt{LatticeComplex result} and perform the following computation:
\begin{verbatim}
    result =  3.0 * map.at(std::make_tuple(0, 0, 0, 0, 1, 1))
            - 3.0 * map.at(std::make_tuple(0, 0, 0, 1, 1, 0));
\end{verbatim}
The factors \texttt{3.0} and \texttt{-3.0} correspond to the product of the $\lambda$-coefficients in \cref{eq:terms}, \texttt{map} is the above mentioned lookup table which uses keys in the form of a tuple containing the six spinor indices identifying the unique term needed for the contraction.
The flavor indices are not mentioned here, since the correlators are already split up according to the flavor index combination, for each of which a set of such functions is defined.

\section{Conclusion and Outlook} \label{sec:outlook}

We have presented our calculational framework for the inclusion of isospin-breaking effects for baryon correlators using isosymmetric CLS ensembles, for which we employ a perturbative method \cite{deDivitiis:2011eh,deDivitiis:2013xla}.
For the baryon operators, we use a construction based on parity and symmetries in flavor and spinor indices \cite{Basak:2005ir} which gives rise to several correlator matrices for the different baryons.
We have implemented several optimizations to deal with the vast amount
of correlators to be computed.

We have tested our code on an ensemble (A654) of size $48\times24^3$ with antiperiodic temporal boundary conditions. 
Preliminary results for the single nucleon correlator suggest that we expect a statistical uncertainty below 1\%.

Over the course of the next months, we intend to perform a spectroscopic analysis of the different correlators we have for A654 including the use of the GEVP method.
Moreover, we are going to generate correlator data for larger $128\times64^3$ ensembles, namely D450 and D452, to perform similar analyses on these ensembles.

\acknowledgments
We thank Ben H\"orz for contributions during the early stages of this work.
ADH is supported by: (i) The U.S. Department of Energy, Office of Science,
Office of Nuclear Physics through the Contract No. DE-SC0012704 (S.M.);
(ii) The U.S. Department of Energy, Office of Science, Office of Nuclear
Physics and Office of Advanced Scientific Computing Research, within the
framework of Scientific Discovery through Advance Computing (SciDAC)
award Computing the Properties of Matter with Leadership Computing Resources.

\bibliographystyle{JHEP}
\bibliography{bibliography.bib}

\providecommand{\href}[2]{#2}\begingroup\raggedright\begin{thebibliography}{10}

\bibitem{deDivitiis:2011eh}
G.M.~de~Divitiis et~al., \emph{{Isospin-breaking effects due to the up-down
  mass difference in Lattice QCD}},
  \href{https://doi.org/10.1007/JHEP04(2012)124}{\emph{JHEP} {\bfseries 04}
  (2012) 124} [\href{https://arxiv.org/abs/1110.6294}{{\ttfamily 1110.6294}}].

\bibitem{deDivitiis:2013xla}
{\scshape RM123} collaboration, \emph{{Leading isospin-breaking effects on the
  lattice}}, \href{https://doi.org/10.1103/PhysRevD.87.114505}{\emph{Phys. Rev.
  D} {\bfseries 87} (2013) 114505}
  [\href{https://arxiv.org/abs/1303.4896}{{\ttfamily 1303.4896}}].

\bibitem{Bruno:2016plf}
M.~Bruno, T.~Korzec and S.~Schaefer, \emph{{Setting the scale for the CLS $2 +
  1$ flavor ensembles}},
  \href{https://doi.org/10.1103/PhysRevD.95.074504}{\emph{Phys. Rev. D}
  {\bfseries 95} (2017) 074504}
  [\href{https://arxiv.org/abs/1608.08900}{{\ttfamily 1608.08900}}].

\bibitem{Carrasco:2015xwa}
N.~Carrasco, V.~Lubicz, G.~Martinelli, C.T.~Sachrajda, N.~Tantalo, C.~Tarantino
  et~al., \emph{{QED Corrections to Hadronic Processes in Lattice QCD}},
  \href{https://doi.org/10.1103/PhysRevD.91.074506}{\emph{Phys. Rev. D}
  {\bfseries 91} (2015) 074506}
  [\href{https://arxiv.org/abs/1502.00257}{{\ttfamily 1502.00257}}].

\bibitem{Bruno:2014jqa}
M.~Bruno et~al., \emph{{Simulation of QCD with N$_{f} =$ 2 $+$ 1 flavors of
  non-perturbatively improved Wilson fermions}},
  \href{https://doi.org/10.1007/JHEP02(2015)043}{\emph{JHEP} {\bfseries 02}
  (2015) 043} [\href{https://arxiv.org/abs/1411.3982}{{\ttfamily 1411.3982}}].

\bibitem{Hayakawa:2008an}
M.~Hayakawa and S.~Uno, \emph{{QED in finite volume and finite size scaling
  effect on electromagnetic properties of hadrons}},
  \href{https://doi.org/10.1143/PTP.120.413}{\emph{Prog. Theor. Phys.}
  {\bfseries 120} (2008) 413}
  [\href{https://arxiv.org/abs/0804.2044}{{\ttfamily 0804.2044}}].

\bibitem{Risch:2017xxe}
A.~Risch and H.~Wittig, \emph{{Towards leading isospin-breaking effects in
  mesonic masses with $O(a)$ improved Wilson fermions}},
  \href{https://doi.org/10.1051/epjconf/201817514019}{\emph{EPJ Web Conf.}
  {\bfseries 175} (2018) 14019}
  [\href{https://arxiv.org/abs/1710.06801}{{\ttfamily 1710.06801}}].

\bibitem{Risch:2018ozp}
A.~Risch and H.~Wittig, \emph{{Towards leading isospin-breaking effects in
  mesonic masses with open boundaries}},
  \href{https://doi.org/10.22323/1.334.0059}{\emph{PoS} {\bfseries LATTICE2018}
  (2018) 059} [\href{https://arxiv.org/abs/1811.00895}{{\ttfamily
  1811.00895}}].

\bibitem{Basak:2005ir}
{\scshape Lattice Hadron Physics (LHPC)} collaboration, \emph{{Clebsch-Gordan
  construction of lattice interpolating fields for excited baryons}},
  \href{https://doi.org/10.1103/PhysRevD.72.074501}{\emph{Phys. Rev. D}
  {\bfseries 72} (2005) 074501}
  [\href{https://arxiv.org/abs/hep-lat/0508018}{{\ttfamily hep-lat/0508018}}].

\bibitem{Blossier:2008tx}
B.~Blossier, G.~von Hippel, T.~Mendes, R.~Sommer and M.~Della~Morte,
  \emph{{Efficient use of the Generalized Eigenvalue Problem}},
  \href{https://doi.org/10.22323/1.066.0135}{\emph{PoS} {\bfseries LATTICE2008}
  (2008) 135} [\href{https://arxiv.org/abs/0808.1017}{{\ttfamily 0808.1017}}].

\bibitem{Blossier:2009kd}
B.~Blossier, M.~Della~Morte, G.~von Hippel, T.~Mendes and R.~Sommer, \emph{{On
  the generalized eigenvalue method for energies and matrix elements in lattice
  field theory}},
  \href{https://doi.org/10.1088/1126-6708/2009/04/094}{\emph{JHEP} {\bfseries
  04} (2009) 094} [\href{https://arxiv.org/abs/0902.1265}{{\ttfamily
  0902.1265}}].

\bibitem{Shintani:2014vja}
E.~Shintani, R.~Arthur, T.~Blum, T.~Izubuchi, C.~Jung and C.~Lehner,
  \emph{{Covariant approximation averaging}},
  \href{https://doi.org/10.1103/PhysRevD.91.114511}{\emph{Phys. Rev. D}
  {\bfseries 91} (2015) 114511}
  [\href{https://arxiv.org/abs/1402.0244}{{\ttfamily 1402.0244}}].

\end{thebibliography}\endgroup

\end{document}